\documentclass[aps,prl,twocolumn,showpacs]{revtex4}
\usepackage{amsmath}
\usepackage{graphicx}
\usepackage{units}  
\usepackage{xspace}
\usepackage{subfigure}
\usepackage{hyperref}
\newcommand{\beq}{\begin{equation}}
\newcommand{\eeq}{\end{equation}}
\newcommand{\bea}{\begin{eqnarray}}
\newcommand{\eea}{\end{eqnarray}}

\newcommand{\sx}{\sigma_{ x}}
\newcommand{\sy}{\sigma_{ y}}
\newcommand{\sz}{\sigma_{ z}}
\newcommand{\LLo}{\ensuremath{\textrm{LL}_0}\xspace}
\newcommand{\moire}{moir\'{e}\xspace}
\newcommand{\Moire}{Moir\'{e}\xspace}
\newcommand{\vf}{v}

\usepackage[usenames]{color}

\begin{document}
\bibliographystyle{apsrev}
 
\title{ Local sublattice-symmetry breaking in rotationally faulted multilayer graphene }
\author{M. Kindermann}
\author{P. N. First}
\affiliation{ School of Physics, Georgia Institute of Technology, Atlanta, Georgia 30332, USA }

\date{\today}
\begin{abstract} 
Interlayer coupling in rotationally faulted graphene multilayers breaks the local sublattice-symmetry of the individual layers. We present a  theory of this mechanism, which  reduces to an effective Dirac model with space-dependent mass  in an important limit. It thus makes a wealth of existing knowledge available for the study of rotationally faulted graphene multilayers.  We demonstrate quantitative agreement between our theory and a   recent experiment.  %Our real-space approach will be useful for investigations of the unique spatial   structure of this system. %The theory holds promise of many other exotic effects.
\end{abstract}

\maketitle 

Experiments indicate that the 10--100 individual graphene layers grown on the carbon-terminated face of SiC are surprisingly well decoupled from one another electronically. Early spectroscopic measurements \cite{sadowski:prl06,orlita:prl08}  found a linear low-energy electronic dispersion to the experimental precision, like that of single-layer graphene \cite{novoselov:sci04,zhang:nat05}. In scanning tunneling microscopy/spectroscopy (STM/STS) measurements the Landau level quantization of the material in a magnetic field was found to be essentially that of single-layer graphene %and distinctly different from conventional 2D electron gases with a quadratic dispersion relation 
 \cite{miller:sci09}. Theoretically it has been shown that this approximate decoupling of different layers is due to a relative twist of the layers with respect to each other \cite{latil:prb07,lopes:prl07,hass:prl08,shallcross:prl08,shallcross:prb10,laissardiere:nal10,mele:prb10}. % Heuristically, the crystal growth process favors twist angles $\theta$ within near $30^\circ$ and within $\approx 4^\circ$ of layer alignment \cite{hass:prl08}.% For the  twist angles that we address in what follows, the interlayer coupling at the dominant wavevectors is perturbative \cite{lopes:prl07}. For those typical rotation angles 
 A renormalization of the electron velocity \cite{lopes:prl07,laissardiere:nal10}, van Hove singularities \cite{andrei:nap10}, and interlayer transport \cite{bistritzer:prb10} have been discussed as residual effects of the interlayer coupling. 
 
\begin{figure}[h]
\includegraphics[width=0.85\linewidth]{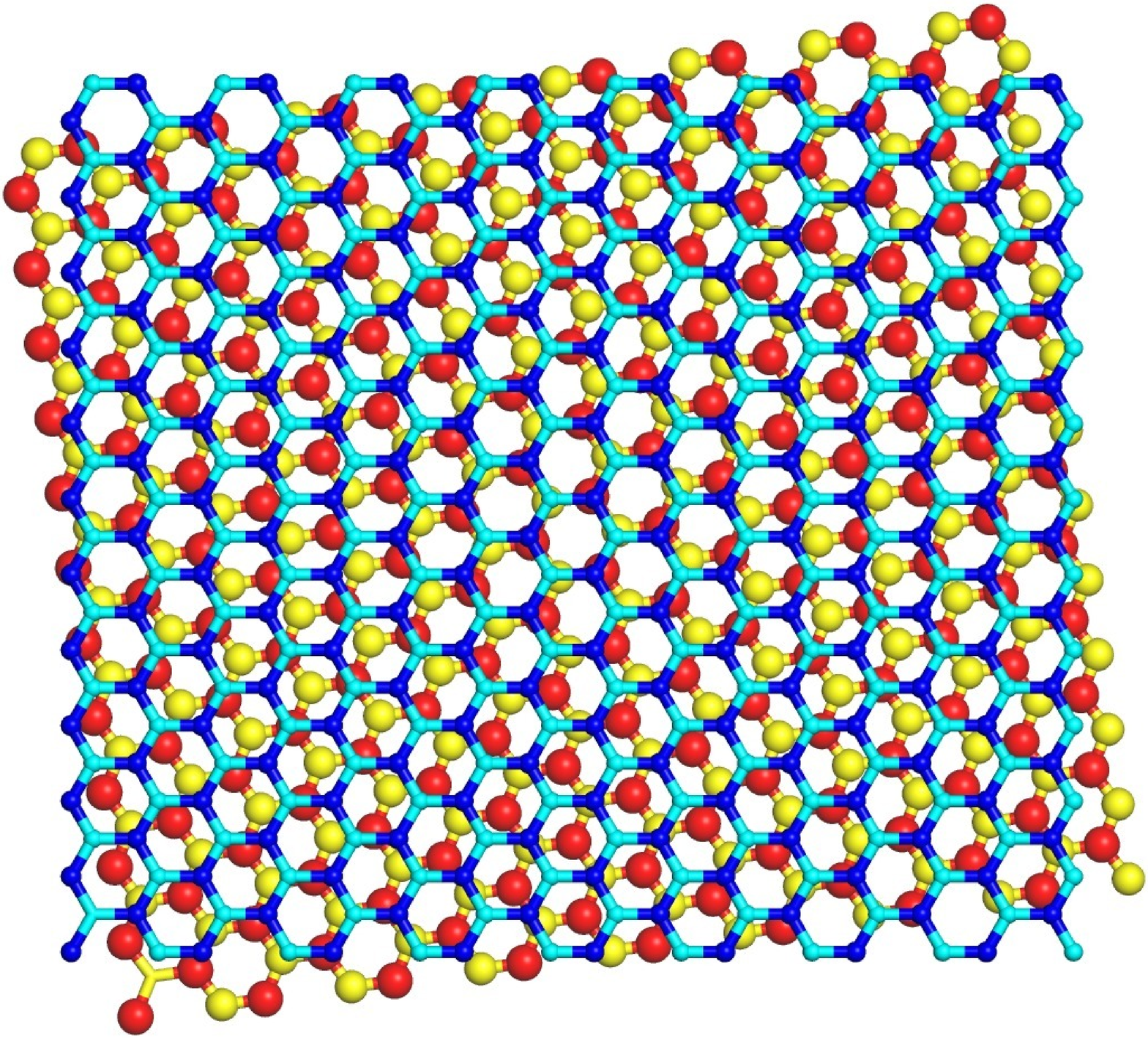}
\caption{(color online) \Moire pattern created by two graphene lattices with a relative twist. Top layer A/B sublattice atoms are shown as small blue/cyan (dark/light) spheres and connectors; bottom layer A/B atoms are shown as large red/yellow (dark/light) spheres. A region of AA alignment lies at the center, where each top-layer atom has a neighbor in the bottom layer. The AA region is surrounded by three AB- and three BA-aligned regions where atoms on only one top-layer sublattice have direct neighbors in the bottom layer. As a consequence, the sublattice-symmetry is broken locally. } \label{fig1} 
\end{figure}

In a recent STM measurement on multilayer epitaxial graphene \cite{miller:nap10} a spatially modulated splitting $\Delta\lesssim \unit[10]{meV}$ of the zeroth Landau level (\LLo) was observed. In view of the above this finding is intriguing, since the states forming \LLo of an isolated layer of graphene without electron-electron interactions are degenerate. % and those interactions are expected to be screened at by the highly doped graphene layers close to the SiC substrate. 
 Therefore, either the observed splitting of \LLo is due to electron-electron interactions, or the interlayer coupling manifests itself prominently in this measurement. In many ways the experimental data favors the latter scenario. One such indication is the observation of a sublattice polarization of the split \LLo: there are regions of space where the branch of \LLo that has positive energy $ \Delta/2 $ appears to consist of wavefunctions localized on the A-sublattice of the graphene lattice, while the lower branch, at negative energy $- \Delta/2$, is localized on the B-sublattice.  The implied local sublattice-symmetry breaking has a natural explanation in terms of the interlayer coupling: the coupling to lower graphene layers generically induces a difference between the local environments of the two sublattices of the top layer in the material, which is probed in STM. This is illustrated in Fig.~\ref{fig1} for a stack of two graphene layers with a relative twist. There are regions where atoms on the A-sublattice of the top layer are closer to atoms in the bottom layer than those on the B-sublattice of the top layer   and regions with the reverse situation. A second conspicuous feature of the STM data is a spatial modulation of the splitting of \LLo: the regions where \LLo is split appear to be arranged on a hexagonal superlattice with a lattice constant $l\approx \unit[70]{nm}$.   It thus shares the symmetries of the \moire pattern characteristic of the twisted graphene bilayer shown in Fig.~\ref{fig1}---another strong indication that the observed splitting is due to the interlayer coupling. 

Earlier theory of the interlayer coupling in graphene multilayers did not predict the observed splitting of \LLo. In Ref.~\cite{miller:nap10} we therefore proposed a phenomenological theory, modeling the different local environments of the A- and the B-atoms of the top graphene layer by a ``staggered'' electric potential $V_{AB}$ that has opposite sign on the two sublattices. This model qualitatively accounts for the main features of the experimental data.  In this Letter we present a microscopic theory of the interlayer coupling in rotationally faulted graphene multilayers. We reduce the problem to an effective model of the  top layer of the material, which is probed in many experiments, such as STM.  In order to conveniently explain the rich spatial structure of the system illustrated in Fig.\ \ref{fig1} and observed in Ref.\ \cite{miller:nap10}  we formulate our theory in real space, as distinct from prior momentum-space approaches  \cite{lopes:prl07,laissardiere:nal10,bistritzer:prb10}. The resulting Hamiltonian reduces to the phenomenological model of Ref.~\cite{miller:nap10} in certain limits and it likewise reproduces the main qualitative features of the measurements. Our theory moreover allows us to test quantitatively whether the interlayer coupling can explain the experimental findings \cite{miller:nap10}. The answer is affirmative: using the commonly accepted tight-binding parameters of graphene multilayers our theory predicts both the magnitude of the observed splitting and its magnetic field dependence in very good agreement with experiment. 
 
 We analyze the electron dynamics in a graphene layer ``$0$'' when coupled to a second layer ``$1$,'' twisted by a relative angle $\theta$ ($\theta=0^\circ$ for aligned honeycomb lattices, \emph{cf.}\ Fig.~\ref{fig1}), neglecting electron-electron interactions. The corresponding dynamics in multilayers at perturbatively weak interlayer coupling, such as in the experiment \cite{miller:nap10}, are obtained by summation over all layers coupled to the top layer $0$. %, where this allows us to model the interaction between any pair of layers. 
 Twisted graphene bilayers have been described before \cite{lopes:prl07,shallcross:prl08,shallcross:prb10,laissardiere:nal10,mele:prb10} by a  local interlayer coupling Hamiltonian with parameters fitted to experiment \cite{dresselhaus:aip02},
   \beq
 {H}_{\rm int}= \int {d\boldsymbol{r} \, \Psi^{(0)\dag}(\boldsymbol{r})\Gamma(\boldsymbol{r})\Psi^{(1)}(\boldsymbol{r})+h.c.}
 \eeq
 Here, the spinors $\Psi^{(j)}$  collect the amplitudes for electrons on the two sublattices of layer $j\in\{0,1\}$.  The interlayer coupling  $\Gamma$ has contributions at wavevectors $\boldsymbol{b}^{(0)}-\boldsymbol{b}^{(1)}$, where $\boldsymbol{b}^{(j)}$ are reciprocal  vectors of the graphene lattice in layer $j$ \cite{mele:prb10}.  The Fourier components   of $\Gamma$ quickly decay with increasing wavevector \cite{mele:prb10,shallcross:prl08,shallcross:prb10}.  %A monotonically decreasing function $h(\delta \boldsymbol{r})$ then describes coupling between sites in different layers that decays with increasing distance $ |\delta \boldsymbol{r}|$ between those sites. 
   In this Letter we therefore neglect all but  the zero wavevector component, setting $\Gamma ({\bf r})=\gamma$, such that the distinction between commensurate and incommensurate interlayer rotations disappears. %We therefore do not make that distinction.
 This approximation is valid for energies $\varepsilon\gg {\cal V}$, where ${\cal V}$ is set by the Fourier components of $\Gamma$ that directly connect K-points of the two layers \cite{mele:prb10}. We take the limit $0<\theta \ll 1$, when    ${\cal V}\ll\ \gamma$ (in the experiment \cite{miller:nap10} $\theta\approx .25^\circ$ and according to the estimate ${\cal V}\simeq \theta^2\gamma$ of Ref.\  \cite{mele:prb10}  this approximation is justified at all accessed energies).
 
In our limit $0<\theta\ll 1$ a long-wavelength description is appropriate, where the isolated layers $j$ are described by Dirac model Hamiltonians (we set $\hbar =1$)
\beq \label{Dirac} 
 H^{(j)}= \vf \int {d\boldsymbol{r}\sum_\nu \psi_\nu^{(j)\dag}(\boldsymbol{r})\left[
\boldsymbol{\sigma_\nu}\cdot \left(-i
 \boldsymbol{\nabla}+e{\bf A}({\bf r})\right)\right]\psi_\nu^{(j)} (\boldsymbol{r})}.
 \eeq
 Here, $\boldsymbol{\sigma_\nu}=(\nu\sx,\sy)$ is a vector of Pauli matrices, $\nu=\pm $ is the valley spin, $-e$ the electron charge, and $\vf$ the electron velocity
in graphene. We have included an external vector potential ${\bf A}$ to describe a perpendicular magnetic field $B$.
Eq.\ (\ref{Dirac}) acts on the long-wavelength spinors $\psi_{\mu,\nu}^{(j)}$ defined by $\Psi^{(j)}_{\mu}(\boldsymbol{r})=\sum_\nu u_{\mu,\nu}^{(j)}(\boldsymbol{r})\psi_{\mu,\nu}^{(j)}(\boldsymbol{r})$. We write the Bloch functions $u_{\mu,\nu}^{(j)}(\boldsymbol{r})=\{\sum_p \exp[i \boldsymbol{K}^{(j)}_{p\nu}\cdot(\boldsymbol{r}-\boldsymbol{\tau}_{\mu}^{(j)})]\}/\sqrt{3}$ in the``first star approximation'' appropriate for the interlayer coupling problem \cite{mele:prb10}. Here, $p$ sums over the three equivalent Brillouin zone corners $\boldsymbol{K}^{(j)}_{p\nu}$ that form the Dirac point of valley $\nu$ \cite{mele:prb10} and   $\boldsymbol{\tau}_{\mu}^{(j)}$  gives the position of an atom on sublattice $\mu \in \{{\rm A,B}\}$ within the unit cell  of layer $j$.  In the long-wavelength theory (which neglects inter-valley processes) the interlayer coupling reads
 \beq
 H_{\rm int} = \int {d\boldsymbol{r} \sum_{\nu} \psi^{(0)\dag}_\nu(\boldsymbol{r})t_\nu(\boldsymbol{r})\psi_\nu^{(1)}(\boldsymbol{r})}+h.c.,
 \eeq
with a matrix $t$ whose long-wavelength components have wavevectors $ \delta\boldsymbol{K}_{p\nu} =(R_\theta-1)\boldsymbol{K}^{(0)}_{p\nu}$. Here, $R_\theta$ is a rotation around the $z$-axis by angle $\theta$. Retaining only those long-wavelength parts of $t$ we find
\beq \label{t}
t^{\mu\mu'}_{\nu}\!\!(\boldsymbol{r})\! =\!\frac{\gamma}{3 }\sum_{p}\!e^{i \delta\boldsymbol{K}_{p\nu}\cdot\boldsymbol{r}+i\boldsymbol{K}_{p\nu}\cdot\left(\boldsymbol{\tau}_{\mu}^{(0)}-\boldsymbol{\tau}_{\mu'}^{(1)}\right)},
\eeq
where terms of order $\theta$ are neglected, while terms of order $\theta K r$ are kept as they may grow large.

\begin{figure}[h]
\subfigure[][~$V^{\rm eff}$]{\includegraphics[width=0.4\linewidth]{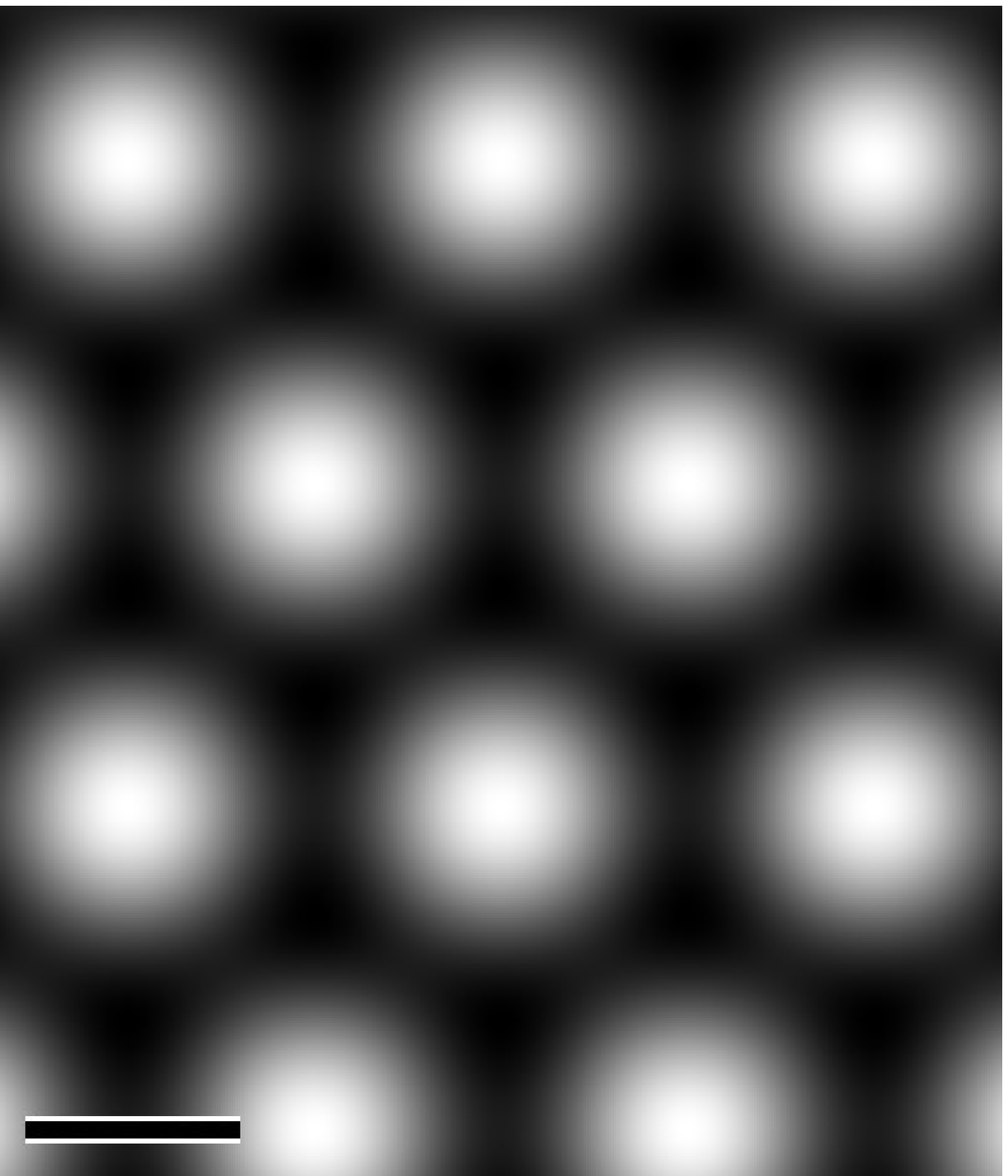}}\hspace{0.08\linewidth}
\subfigure[][~$m^{\rm eff}$]{\includegraphics[width=0.4\linewidth]{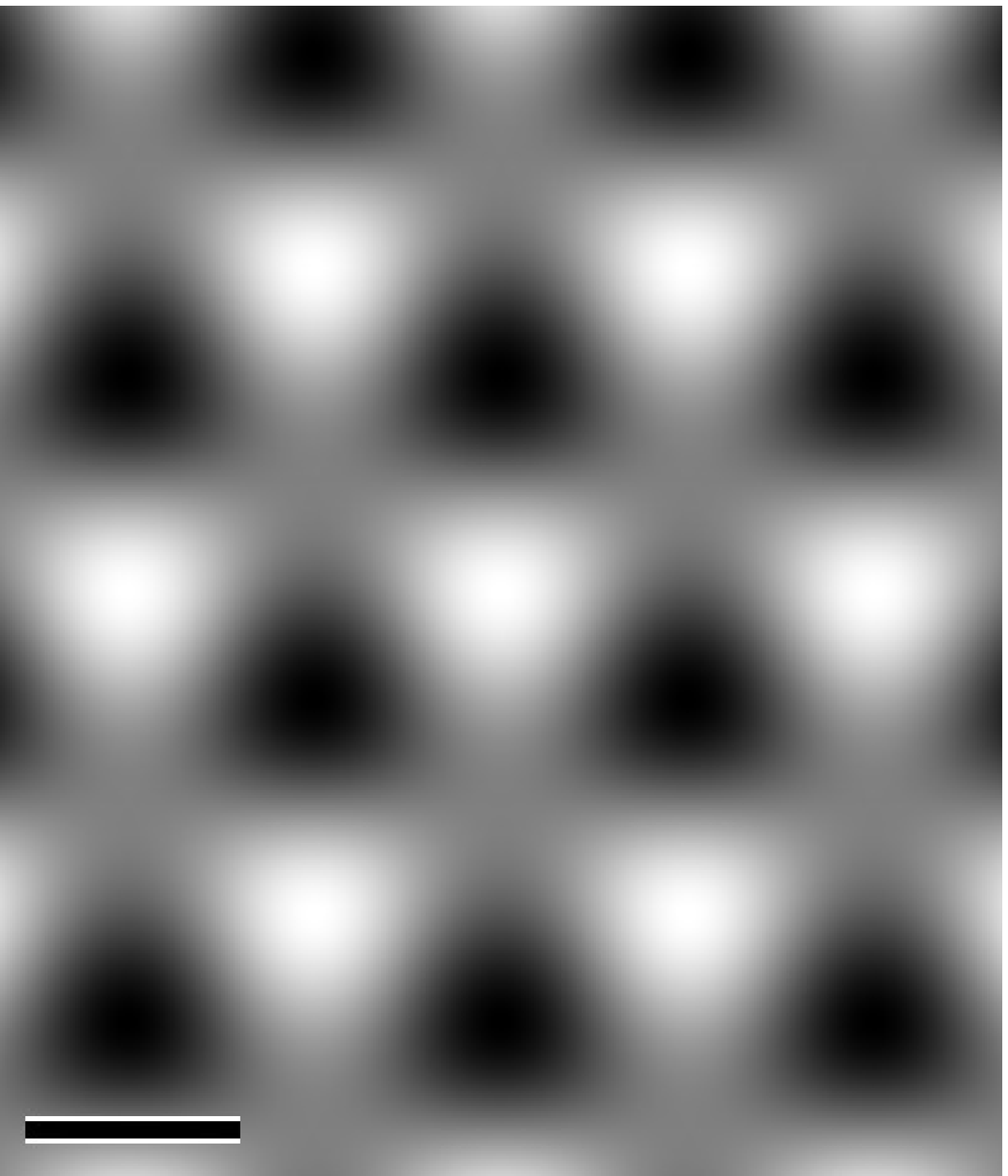}}
\subfigure[][~$A_x^{\rm eff}$]{\includegraphics[width=0.4\linewidth]{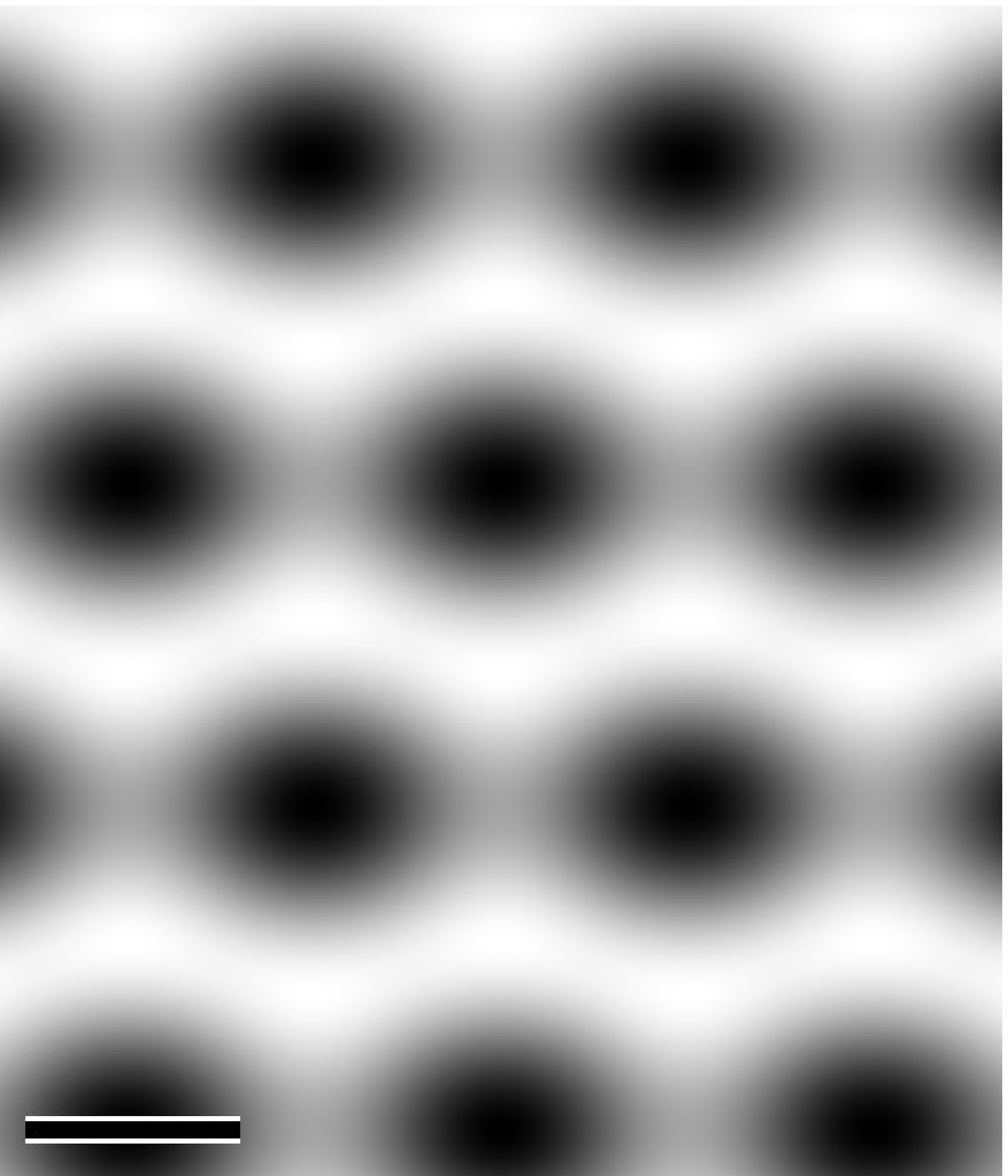}}\hspace{0.08\linewidth}
\subfigure[][~$A_y^{\rm eff}$]{\includegraphics[width=0.4\linewidth]{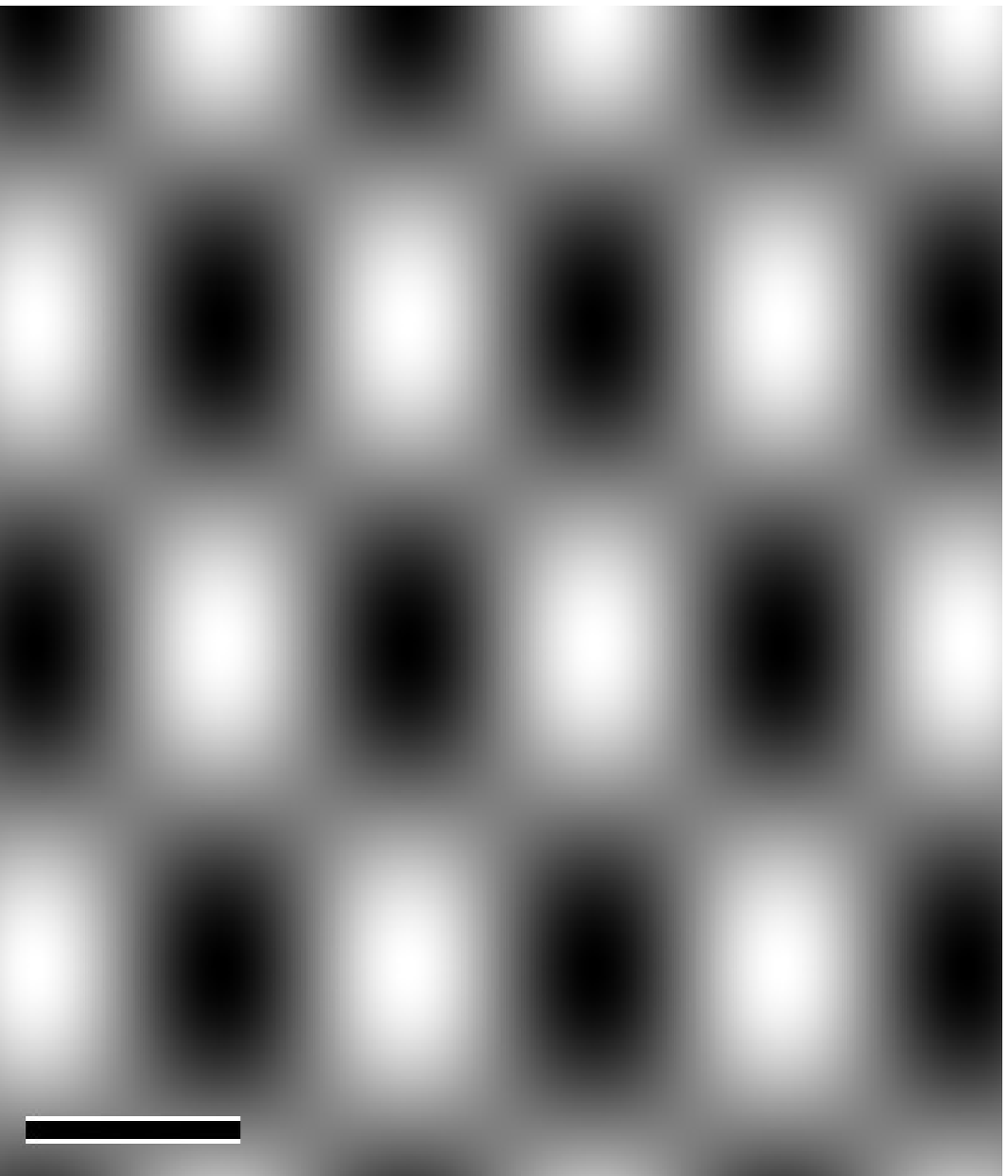}}
\caption{ (a) Effective potential $V^{\rm eff}$, (b) effective mass $m^{\rm eff}$, (c) $A_x^{\rm eff}$, and (d) $A_y^{\rm eff}$ of Eq.\ (\ref{para}) as functions of ${\bf r}\theta/a$ in grey-scale. Scale bars span a unity increment in ${\bf r}\theta/a$. Note the expected sixfold and threefold symmetries of $V^{\rm eff}$ and $m^{\rm eff}$, respectively. ${\bf A}^{\rm eff}$ transforms as a vector under rotations. } \label{fig2} 
\end{figure}

We next integrate out layer $j=1$ in order to arrive at an effective Hamiltonian $H_0^{\rm eff}(\omega)=H_0+\delta H_0^{\rm eff}(\omega)$ for the top layer $j=0$, with
\beq \label{Heff}
\delta H_0^{\rm eff}(\omega)= H_{\rm int}(\omega+V-H_1)^{-1}H_{\rm int}.
\eeq
We include an interlayer bias $V$ that accounts for different doping levels of the two layers \footnote{In the generalization to multilayers the interaction between layers with $j>0$ needs to be added to the diagonal part $\omega-\sum_j H_j$. }.
In general, $H_0^{\rm eff}$ is nonlocal in space and it depends on the energy $\omega$. In the limit of a large interlayer bias, however, $|V|\gg \omega, \gamma, \theta v/a $, the sum $\omega+V-H_1$ becomes momentum- and energy-independent to a good approximation. The spatial nonlocality and the energy-dependence of $H_0^{\rm eff}$  then may be neglected and $H_0^{\rm eff}$ becomes a conventional Dirac Hamiltonian (\ref{Dirac}) with a matrix potential 
\beq \label{local}
\delta H_{0}^{\rm eff}= \int {d\boldsymbol{r}\sum_\nu \psi_\nu^{(0)\dag} (\boldsymbol{r})\frac{t_\nu({\bf r})t_\nu^\dag({\bf r})}{V} \psi_\nu^{(0)} (\boldsymbol{r})},
\eeq
which we parametrize as
\beq \label{para}
\frac{t_\nu({\bf r})t_\nu^\dag({\bf r})}{V}= V^{\rm eff}({\bf r})+\nu \vf e\boldsymbol{\sigma}_\nu \cdot \boldsymbol{A}^{\rm eff}({\bf r})+m^{\rm eff}({\bf r}) \vf^2 \sz.
\eeq
%In our approximation ${\bf b}=0$ in Eq.\ (\ref{t}) we find from Eq.\ (\ref{local})
%\beq
%\delta H_{0\nu; \mu\mu'}^{\rm eff}({\bf r})=-\frac{|\gamma|^2}{9V \Omega^4}\sum_{p, q} e^{i(\delta{\bf K}_p-\delta{\bf K}_q)\cdot {\bf r}+i{\bf K}_p
%\eeq
The interlayer coupling in this limit generates effective scalar and vector potentials $V^{\rm eff}$ and ${\bf A}^{\rm eff}$, respectively, and a mass term $\propto \sz m^{\rm eff} \vf^2$ that implies an effective staggered potential $V^{\rm eff}_{AB}=m^{\rm eff}\vf^2$ in locally Bernal stacked regions.
It follows from Eq.\ (\ref{t}) that $\delta H_{0\nu}^{\rm eff}$ oscillates in space with wavevectors ${\bf k}=(R_\theta-1)\boldsymbol{b}$, where ${\bf b}$ is in the ``first star'' of reciprocal lattice vectors of graphene. 
We plot $\delta H_{0\nu}^{\rm eff}$ in the parameterization of Eq.\ (\ref{para}) in Fig.\ \ref{fig2}. 

Now turning to the experiment \cite{miller:nap10} we note that at large interlayer bias $V$ our theory  takes the form of the phenomenological Hamiltonian proposed in Ref.\ \cite{miller:nap10}. It then intuitively explains the main qualitative features of the experiment: perturbatively in $\gamma$, the energy shift of a \LLo wavefunction $\psi_{0,\nu}$ in valley $\nu$ is given by 
\beq \label{split}
\delta\varepsilon_{0,\nu}= \langle \psi_{0\nu}|\delta H^{\rm eff}_{0\nu}(\omega=0)|\psi_{0\nu}\rangle.
\eeq
The unperturbed  \LLo wavefunctions are localized on individual sublattices. Therefore, if $\delta H^{\rm eff}_0$ included a constant staggered potential $V_{AB}>0$, with potentials $ V_{AB}$ and $-V_{AB}$ for atoms on the A- and B-sublattice, respectively, a splitting $\Delta=\delta\varepsilon_{0,\nu=1}-\delta\varepsilon_{0,\nu=-1}=-2V_{AB}$  between sublattice-polarized \LLo states would result, as observed experimentally: $ V_{AB}$ would  increase the energy of the states $\psi_{0,\nu=-1}$ localized on the A-sublattice and decrease the energy of the $\nu=1$ states, localized on the B-sublattice.  For the space-dependent $V^{\rm eff}_{AB}=m^{\rm eff}\vf^2$ of Fig.\ \ref{fig2} that splitting is still present  locally, around the extrema of $m^{\rm eff}$, at sufficiently large magnetic fields $B$, when the \LLo wavefunctions fit   well into the regions with extremal $m^{\rm eff}$. Comparison of Fig. \ref{fig2} with Fig.\ 5a of Ref.\ \cite{miller:nap10} shows that the thus predicted spatial symmetries of $|\Delta|\propto |m^{\rm eff}|$ agree with experiment. For large $B$  the splitting approaches $\lim_{B\to \infty} \Delta=-2V^{\rm eff}_{AB}$. With decreasing $B$, as the wavefunctions become more extended, $\Delta$ gets averaged over maxima and minima of $m^{\rm eff}$ and it is suppressed, also in accordance with experiment.% Higher Landau levels are not sublattice-polarized and $V_{AB}$ has no effect to leading order. Also this is consistent with experiment, where the higher Landau levels where hardly affected in the regions where \LLo is split (see for example Fig.\ 3 of Ref.\ \cite{miller:sci09}).

The experiment of Ref.\ \cite{miller:nap10}, however, was not done in the high bias limit. The fact that in the measurement \cite{miller:nap10} tunneling into \LLo occurred only at a finite bias voltage $V_{\rm STM}\approx \unit[60]{meV}$ between STM-tip and sample does indicate a doping of the graphene layers at the surface. The difference between the chemical potentials of the top layer and the layers below after screening is expected to be $|V|\lesssim V_{\rm STM} \approx\unit[60]{meV} $. However, the large applied magnetic field $ \unit[4]{T} \leq B\leq \unit[8]{T}$   corresponds to a large cyclotron frequency $\omega_c=\sqrt{2}\vf/l_B$ \cite{neto:rmp09}, where $l_B=1/\sqrt{eB}$ is the magnetic length: $\omega_c\approx\unit[105]{meV}$ at $B=\unit[8]{T}$. In this experiment therefore $|V| < \omega_c$ and $H_0^{\rm eff}$ is not local on the scale $l_{\rm B}$ on which the wavefunctions vary. 

The experiment also indicates that it is the coupling between the top layer and its next-to-nearest layer (that is the third layer from the top) that produces the observed splitting. One concludes this from the observation that the dominant \moire of the STM topography, most likely due to the coupling of the top layer to its nearest neighbor, has a much smaller lattice constant $\bar{l}\approx \unit[4]{nm}$ than the superlattice associated with the splitting of \LLo with $l\approx \unit[70]{nm}$. The estimates of the next-to-nearest layer coupling in the literature vary \cite{dresselhaus:aip02,nilsson:prb08,brandt:bo88,chung:jms02}, but there is a consensus that the coupling constant is $\gamma\lesssim \unit[40]{meV}$. The physics at the energies $\omega=\pm \Delta/2\approx \pm \unit[5]{meV}$ where the splitting of \LLo occurs is thus described by $H_0^{\rm eff}$ at $|\omega|,\gamma \ll |V| \ll \omega_c$. In this limit the effects of the interlayer coupling are perturbative, which allows us to deal with the non-locality of $ H_{0}^{\rm eff}$ analytically. We evaluate Eq.\ (\ref{split}) at $|\omega|,\gamma \ll |V| \ll \omega_c$ in the appendix. In accordance with the intuition gained from the limit $V\to \infty$ of the previous paragraph, the resulting $\Delta$ is  extremal in locally Bernal stacked regions and the wavefunctions are sublattice-polarized. The qualitative agreement with experiment thus carries over to the non-local theory.

Now comparing our theory also quantitatively with the experiment we first take the limit of a large magnetic field, when the wavefunctions fit well into the Bernal stacked regions. The maximal splitting $\Delta_{\rm max}$, reached at $B\to\infty$ in AB- or BA-stacked regions, can be extracted from Eq.\ (\ref{deleps}) of the appendix by taking the limit $\theta\to 0$ at fixed $B$. We find
 \beq \label{delmax}
|\Delta_{\rm max}|= |V|\left|\frac{\gamma}{\omega_c}\right|^2 
\eeq
in our approximations. Estimating $\gamma$ by $\gamma=\gamma_5 \approx \unit[38]{meV}$ given in Ref.\ \cite{brandt:bo88} we find that $|\Delta_{\rm max}| \approx \unit[5]{meV}$ for $V\approx \unit[40]{meV}$. Considering the uncertainties in our knowledge of $\gamma$ and $V$, this agrees well with the experimentally observed value $\Delta \approx \unit[10]{meV}$.

We next quantify the magnetic field dependence of $\Delta$ in the regions with maximal $\Delta$ at $B\to \infty$ (that is AB- or BA-stacked regions) by expanding Eq.\ (\ref{deleps}) asymptotically for $\delta Kl_B\gg1$: 
\bea\label{del}
%\delta &\sim &\frac{V}{2} \left|\frac{ 9 \nu a}{4\pi\vf\theta}\right|^2 e^{-(5/3)(4\pi\theta l_B/3a)^2} \nonumber \\
|\Delta| &\sim & \frac{2|V|}{\sqrt{3} }\left|\frac{\gamma }{v\delta K}\right|^2 e^{-B_*/B}  \\
&&\mbox{} \!\!\!\!\!\!\!\!\!\times \left| \cos \left(\frac{\sqrt{3}B_*}{5B} -\frac{\pi}{6}\right)-\frac{2 v\delta K}{V}\cos\left( \frac{\sqrt{3}B_*}{5B}+\frac{\pi}{6}\right)  \right|.\nonumber
\eea
 The crossover field $B_*=5\delta K^2/4e$, where the exponent in Eq.\ (\ref{del}) becomes of order $1$ and $\Delta$ starts to be exponentially suppressed, evaluates to $B_*=\unit[4.8]{T}$ for the interlayer rotation angle $\theta = 0.25^\circ$ of the \moire pattern in the experiment of Ref.\ \cite{miller:nap10}. Also that crossover field compares favorably with the experiment, where the splitting $\Delta$ disappears between $B\approx \unit[4]{T}$ and $B\approx \unit[6]{T}$. Clearly therefore, the interlayer coupling can account for the main features of the splitting of \LLo reported in Ref.\ \cite{miller:nap10} also on a quantitative level. 

We finally discuss the influence of the graphene layers in the experimental sample that we have ignored so far. The coupling of the top layer to layers further away than the third layer from the top is negligibly small. The coupling $\bar{\gamma}$ to the second layer, however, is not: $\bar{\gamma}\approx 0.4 \,{\rm eV}$ \cite{nilsson:prb08}. As mentioned before, the STM topography of Ref.\ \cite{miller:nap10} has a \moire pattern with scale $\bar{l}\approx \unit[4]{nm}$, which indicates a  rotation angle between the top two layers of $\bar{\theta}\approx 4 ^\circ$. At this angle the coupling between the``first stars'' of the Brillouin zones of those two layers is perturbative, because of large energy denominators \cite{lopes:prl07}. The coupling between other $K$-points in the extended Brillouin zone is too small to play a  role at the scale of the observed splitting $\Delta$ \cite{mele:prb10}. The  perturbative calculation outlined in the appendix therefore describes also the coupling between the top two layers of the measured sample. Applying Eq.\ (\ref{del}) to that coupling we find an exponential suppression of $\Delta$ that is lifted only above a crossover field $\bar{B}_*=(\bar{\theta}/\theta)^2 B_*\approx 300 B_* $  that is much larger than the experimentally applied fields. The only interlayer coupling relevant to the experiment of Ref.\ \cite{miller:nap10} is therefore the next-to-nearest layer coupling discussed above.
 
 We conclude that the interlayer coupling is a viable explanation of the splitting of \LLo reported in Ref.\ \cite{miller:nap10}, both qualitatively and quantitatively. The theory that allowed us to reach these conclusions reduces in certain limits to an effective Dirac model for the top layer of a multilayer system, with effective potentials and a space-dependent mass. As such it makes the wealth of knowledge and intuition existing for the physics of single layer graphene available for the study of rotationally faulted multilayer graphene. Our theory thus appears to be an advantageous starting point for the exploration of much of the physics of this rather complex system. Numerous unconventional and so far unexplained phenomena observed in the material \cite{deheer:ssc07} as well as known properties of our theory promise that such exploration will be rewarding. Especially the effective mass term is expected to have profound implications, for instance topologically confined states \cite{martin:prl08,semenoff:prl08}.

\begin{appendix}

\section{Appendix: Perturbative  Landau level splitting in a large magnetic field }\label{appa}
 We evaluate the splitting $\Delta=\delta\varepsilon_{0,\nu=1}-\delta\varepsilon_{0,\nu=-1}$ between the two valleys of \LLo  at $|\omega|,\gamma \ll |V| \ll \omega_c$, when it is perturbative, using Eq.\ (\ref{split})  with localized wavefunctions of \LLo: $\psi_{0,\nu=1}=(0,\exp[-(x^2+y^2)/4l_B^2 + ixy/2l_B^2]/\sqrt{2\pi}l_B)$ and $\psi_{0,\nu=-1}=\sy \psi_{0,\nu=1}$. 
We write the effective Hamiltonian  as
\beq \label{dH}
\delta H^{\rm eff}_\nu(\boldsymbol{r},\boldsymbol{r'},\omega)= t_\nu(\boldsymbol{r})G_\nu(\omega,\boldsymbol{r},\boldsymbol{r'})t_\nu^{\dag}(\boldsymbol{r'}),
\eeq
where
\beq \label{inv}
G_\nu(\omega,\boldsymbol{r},\boldsymbol{r'})= \int \frac{dk}{2\pi}\sum_{n\geq 0,s=\pm} \frac{\psi_{snk\nu}(\boldsymbol{r})\psi^\dag_{snk\nu}(\boldsymbol{r}')}{\omega+V-s \varepsilon_n} 
\eeq
with
\beq \label{psi}
\psi_{snk\nu=1}(\boldsymbol{r})=\frac{1}{\sqrt{2l_B}} \left(\begin{array}{c} \Phi_{n-1}\left(\frac{y}{l_B}-kl_B\right)\\ s\Phi_{n}\left(\frac{y}{l_B}-kl_B\right)\end{array}\right)e^{i kx} \;\; (n\geq0)
\eeq
in terms of the oscillator wavefunctions
\beq \label{osc}
 \Phi_n(\chi)=\frac{(-1)^n}{\sqrt{2^nn!\sqrt{\pi}}} e^{\chi^2/2}\frac{d^n}{d\chi^n}e^{-\chi^2}.
\eeq
Here, $\Phi_{-1}=0$, $\varepsilon_n=\sqrt{n}\omega_c$ and the wavefunctions in the valley $\nu=-1$ are obtained as $\psi_{s,n,k,\nu=-1}=\sy \psi_{snk\nu=1}$.
 In our limit $|V|\ll \omega_c$, the contribution to $\delta\varepsilon_{0,\nu}$ with the smallest energy denominator comes from the term in Eq.\ (\ref{inv}) with $n=0$. That term is $\propto |t_{BB}|^2$ in valley $\nu=1$. In valley $\nu=-1$ the corresponding term is identical, except that $t_{BB}$ is replaced by $t_{AA}$. One has $t_{BB}=t_{AA}+{\cal O}(\theta)$ \footnote{To leading order in $\theta$ the effect of the interlayer rotation is a space-dependent translation of the unit cells in the two layers with respect to each other. To every pair of an $A$-atom in the top layer and an A-atom in the bottom layer there is therefore a pair of B-atoms with the same distance and therefore the same coupling strength.}. To leading order in $\theta$ this term therefore does not contribute to $\Delta$.  The dominant contribution to $\delta\varepsilon_{0}$ thus comes from the off-diagonal elements of $\delta H^{\rm eff}$ and from the diagonal elements  that are $\propto |t_{AB}|^2$ or $\propto |t_{BA}|^2$  [the upper diagonal element in Eq.\ (\ref{inv}) at $\nu=1$]. In those matrix elements all contributing energy denominators are of the same order, ${\cal O}(\omega_c)$. %In the limit $\theta l_B/a \gg 1 $, when the interlayer coupling creates matrix elements between different Landau levels, 
We thus need to carry out the sum over $n$ in Eq.\ (\ref{inv}). We do this below for $G_1$. The Green function in the other valley is then obtained as $G_{-1}=\sy G_1\sy$. We first rewrite Eqs.\ (\ref{inv}) with (\ref{psi}) and (\ref{osc}) as 
\bea \label{gdef}
G_{1}(\omega,\boldsymbol{r},\boldsymbol{r'})\!&=&\!\!\int\! \frac{dk}{2\pi}\,e^{ik(x-x')+[(y/l_B-kl_B)^2+(y'/l_B-kl_B)^2]/2} \nonumber \\
&&\times g\left(\omega,\frac{y}{l_B}-kl_B,\frac{y'}{l_B}-kl_B\right)
 \eea 
 and note that in our limit $|\omega|\ll |V| \ll \omega_c$ the component $g_{AA}$, which makes one of the leading contributions to $\Delta$ according to the above considerations,  can be expressed as 
\beq \label{diff}
 g_{AA}(0,\chi,\chi')=\frac{ \sqrt{\pi}V}{2\omega_c^2} l(\chi,\chi') +{\cal O}\left(\frac{V^2}{\omega_c^3}\right) 
 \eeq
in terms of a function $l$ that solves the differential equation
\beq \label{dl}
\frac{\partial^2l}{\partial \chi\partial \chi'}=-\frac{4}{ \sqrt{\pi}}\sum_{n\geq 1}\Phi_{n}(\chi)\Phi_{n}(\chi') e^{-(\chi^2+\chi'^2)/2} .
\eeq
 Using the completeness of the oscillator wavefunctions we find that Eq.\ (\ref{dl}) is solved by 
\bea\label{l}
l(\chi,\chi')&=&\left[{\rm erf}(\chi)-{\rm sgn}\,(\chi-\chi')\right]\left[{\rm erf}(\chi)+{\rm sgn}\,(\chi-\chi')\right] \nonumber \\
&& +f(\chi)+f'(\chi')
\eea
%such that
%\begin{widetext}
%\beq
%g(\chi,\chi')=\left(\begin{array}{cc} \frac{2\sqrt{\pi}V}{\omega_c^2} & -\frac{\sqrt{2\pi}}{\omega_c} \frac{\partial}{\partial \chi'}\\-\frac{\sqrt{2\pi}}{\omega_c} \frac{\partial}{\partial \chi}& 0\end{array}\right)l(\chi,\chi')+{\cal O}\left(\frac{V}{\omega_c}\right)^2-\left(\begin{array}{cc} 0&0\\0& e^{-(\chi^2+\chi'^2)/2}\Phi_{0}(\chi)\Phi_{0}(\chi')/V+{\cal O}\left(\frac{V}{\omega_c}\right)^0\end{array}\right)
%\eeq
%\end{widetext}
 with arbitrary functions $f$ and $f'$. Exploiting the symmetries $g(\omega,\chi',\chi)= g(\omega,\chi,\chi')$ and $g(\omega,-\chi,-\chi')= g(\omega,\chi,\chi')$ that are implied by Eqs.\ (\ref{inv}) and (\ref{gdef}), one finds that $f=f'$ and that $f$ is an odd function of $\chi$. %Additionally, $l$ solves also $(-\partial_\chi^2/2-\chi \partial_\chi)l=(V/\omega_c)\exp(\chi^2) \delta(\chi-\chi')$, which follows from $(\chi^2-\partial_\chi^2-1) \Phi_n=n\Phi_n$ in Eq.\ (\ref{inv}). This implies that $f(\chi) \propto\sqrt{\pi}/2- {\rm erf}\,(\chi)$. 
 Now noting that according to Eq.\ (\ref{inv}) $\langle \Phi_n|G(\omega)|\Phi_m\rangle=0$ for $n\neq m$ one concludes that $f=0$ \footnote{This is seen easiest when $n+ m$ is odd by the transformation $(\chi,\chi')\to(-\chi,-\chi')$ in the expression for $\langle \Phi_n|l(\chi,\chi')\exp(\chi^2/2+\chi'^2/2)|\Phi_m\rangle$.}. 
 The off-diagonal matrix elements of $g$ are  found similarly.  To leading order in $V$ they read
 \bea
g_{AB}(0,\chi,\chi')&=&\frac{\sqrt{2\pi}}{4\omega_c} \frac{\partial}{\partial \chi'}l(\chi,\chi')+{\cal O}\left(\frac{V}{\omega_c^2}\right),\nonumber \\
g_{BA}(0,\chi,\chi')&=& g_{\rm AB}(0,\chi',\chi).
\eea
Eqs.\ (\ref{dH}), (\ref{gdef}), (\ref{diff}) and (\ref{l})   allow us to evaluate $\delta\varepsilon_{0,\nu=1}$, Eq.\ (\ref{split}), to leading order in $\gamma$, yielding 
\begin{widetext}
\bea \label{deleps}
\delta\varepsilon_{0,\nu=1}&=& \frac{V}{2\omega_c^2}\int d\chi d\chi'\sum_{ p,p'} \left[{\rm erf}(\chi)-{\rm sgn}(\chi-\chi')\right] \left[{\rm erf}(\chi)+{\rm sgn}(\chi-\chi')\right] \nonumber \\
&&\mbox{} \!\!\!\!\!\!\! \!\!\!\!\!\!\! \!\!\!\!\!\!\!\times \left\{t^*_{\rm BA}(\boldsymbol{\delta\boldsymbol{K}_{p}})t_{\rm BA}(\boldsymbol{\delta\boldsymbol{K}_{p'}}) +(\omega_c/\sqrt{2}V)\left[ t^*_{\rm BB}(\boldsymbol{\delta\boldsymbol{K}_{p}})t_{\rm BA}(\boldsymbol{\delta\boldsymbol{K}_{p'}})(  {\bf c}\cdot\delta\boldsymbol{K}_{p'})^*+ t^*_{\rm BA}(\boldsymbol{\delta\boldsymbol{K}_{p}})t_{\rm BB}(\boldsymbol{\delta\boldsymbol{K}_{p'}})(  {\bf c}\cdot\delta\boldsymbol{K}_{p})\right]\right\} \nonumber \\
%e^{-[(2(G_y-G_y')^2+(G_y+\delta K_{\nu p y} +G_y'+\delta K_{\nu p' y})^2+(G_\chi-G_\chi')^2+2i (G_y+\delta K_{\nu p y} +G_y'+\delta K_{\nu p' y})(G_\chi-G_\chi')]/4-[i(G_\chi+\delta K_{\nu p \chi})+(G_y+\delta K_{\nu p y})]\chi+[i(G_\chi'+\delta K_{\nu p' \chi})-(G_y'+\delta K_{\nu p' y})]\chi'} \\
&&\mbox{} \!\!\!\!\!\!\! \!\!\!\!\!\!\! \!\!\!\!\!\!\!\times e^{-l_{\rm B}^2[(2(\delta\boldsymbol{K}_{p,x}-\delta\boldsymbol{K}_{p',x})^2+(\delta\boldsymbol{K}_{p,x}+\delta\boldsymbol{K}_{p',x})^2+(\delta\boldsymbol{K}_{p,y}-\delta\boldsymbol{K}_{p',y})^2+2i (\delta\boldsymbol{K}_{p,x}+\delta\boldsymbol{K}_{p',x})(\delta\boldsymbol{K}_{p,y}-\delta\boldsymbol{K}_{p',y})]/4- l_{\rm B}[( {\bf c}\cdot\delta\boldsymbol{K}_{p})\chi+( {\bf c}\cdot\delta\boldsymbol{K}_{p'})^*\chi']} , \nonumber \\
\eea
\end{widetext}
where scalar multiplication with ${\bf c}=(1,i)$ maps a vector ${\bf a}$ onto its counterpart ${\bf c}\cdot{\bf a}$ in the complex plane. Here, all wavevectors $\delta \boldsymbol{K}_p$ are evaluated in valley $\nu=1$. The energy shift $\delta\varepsilon_{0,\nu=-1}$ in the other valley is obtained as in Eq.\ (\ref{deleps}), but with $t$ replaced by $\sy t\sy$ and $\delta \boldsymbol{K}_p$   evaluated in valley $\nu=-1$.
In  Eq.\ (\ref{delmax}) of the main text, that is in the limit of large $B$,  only the first term in the curly brackets of Eq.\ (\ref{deleps}) contributes and the resulting splitting $\Delta$ has extrema in regions where the layers are locally Bernal stacked and $|t_{\rm BA}|^2-|t_{\rm AB}|^2$ is extremal. 
 
In the limit $B\to 0$, when the wavefunctions become more and more extended and start averaging over several unit cells of the 
\moire superlattice, the splitting of \LLo decays to zero. In order to quantify this decay of $\Delta$, Eq.\ (\ref{deleps}) may be expanded asymptotically in a large $\delta Kl_B$. Then $\delta\varepsilon_{0,\nu} $ is dominated by the terms with the weakest exponential decay in $\delta Kl_B$, which give
\begin{widetext}
\bea \label{delta}
%\delta \sim V \sum_{p,p'} \left[ t^*_{\rm BA}(\delta K_{p}) t_{\rm BA}(\delta K_{p'})- t^*_{\rm AB}(\delta K_{p'}) t_{\rm AB}(\delta K_{p})\right] \frac{e^{l_B^2|\delta K_p-\delta K_{p'}|^2/2}\left(1-e^{-l_B^2[\delta K_p\cdot \delta K_{p'} + i (\delta K_p\times \delta K_{p'})\cdot \hat{z}]/2}\right)}{\vf^2(\delta K_{p,x}-i\delta K_{p,y})(\delta K_{p',x}+i\delta K_{p',y})}.
\delta\varepsilon_{0,\nu=1}& \sim& V \sum_{p,p'} \frac{ t^*_{\rm BA}(\boldsymbol{\delta\boldsymbol{K}_{p}})t_{\rm BA}(\boldsymbol{\delta\boldsymbol{K}_{p'}})+ (\omega_c/\sqrt{2}V)\left[ t^*_{\rm BB}(\boldsymbol{\delta\boldsymbol{K}_{p}})t_{\rm BA}(\boldsymbol{\delta\boldsymbol{K}_{p'}})(  {\bf c}\cdot\delta\boldsymbol{K}_{p'})^*+ t^*_{\rm BA}(\boldsymbol{\delta\boldsymbol{K}_{p}})t_{\rm BB}(\boldsymbol{\delta\boldsymbol{K}_{p'}})(  {\bf c}\cdot\delta\boldsymbol{K}_{p})\right]}{\vf^2({\bf c}\cdot\delta {\bf K}_{p})({\bf c}\cdot\delta {\bf K}_{p'})^*} \nonumber \\
&&\mbox{} \;\;\;\;\;\;\;\;\;
\times e^{-l_B^2[(\delta \boldsymbol{K}_p-\delta \boldsymbol{K}_{p'})^2+\delta \boldsymbol{K}_p\cdot \delta \boldsymbol{K}_{p'} - i (\delta \boldsymbol{K}_p\times \delta \boldsymbol{K}_{p'})\cdot \hat{z}]/2}
\eea
\end{widetext}
at $\delta Kl_B\gg 1$. Here, $\hat{z}$ is the unit vector along the $z$-axis.
Again all wavevectors $\delta \boldsymbol{K}_p$ are evaluated in valley $\nu=1$ and  $\delta\varepsilon_{0,\nu=-1}$ in the other valley is obtained by replacing $t$ with $\sy t\sy$ in Eq.\ (\ref{delta}) and evaluating  $\delta \boldsymbol{K}_p$    in valley $\nu=-1$.     The sum over $p$ and $p'$ in Eq.\ (\ref{delta}) results in Eq.\ (\ref{del}) of the main text.

\end{appendix}


\begin{thebibliography}{21}
\expandafter\ifx\csname natexlab\endcsname\relax\def\natexlab#1{#1}\fi
\expandafter\ifx\csname bibnamefont\endcsname\relax
  \def\bibnamefont#1{#1}\fi
\expandafter\ifx\csname bibfnamefont\endcsname\relax
  \def\bibfnamefont#1{#1}\fi
\expandafter\ifx\csname citenamefont\endcsname\relax
  \def\citenamefont#1{#1}\fi
\expandafter\ifx\csname url\endcsname\relax
  \def\url#1{\texttt{#1}}\fi
\expandafter\ifx\csname urlprefix\endcsname\relax\def\urlprefix{URL }\fi
\providecommand{\bibinfo}[2]{#2}
\providecommand{\eprint}[2][]{\url{#2}}

\bibitem[{\citenamefont{Sadowski et~al.}(2006)\citenamefont{Sadowski, Martinez,
  Potemski, Berger, and de~Heer}}]{sadowski:prl06}
\bibinfo{author}{\bibfnamefont{M.~L.} \bibnamefont{Sadowski}},
  \bibinfo{author}{\bibfnamefont{G.}~\bibnamefont{Martinez}},
  \bibinfo{author}{\bibfnamefont{M.}~\bibnamefont{Potemski}},
  \bibinfo{author}{\bibfnamefont{C.}~\bibnamefont{Berger}}, \bibnamefont{and}
  \bibinfo{author}{\bibfnamefont{W.~A.} \bibnamefont{de~Heer}},
  \bibinfo{journal}{Phys. Rev. Lett.} \textbf{\bibinfo{volume}{97}},
  \bibinfo{pages}{266405} (\bibinfo{year}{2006}).

\bibitem[{\citenamefont{Orlita et~al.}(2008)\citenamefont{Orlita, Faugeras,
  Plochocka, Neugebauer, Martinez, Maude, Barra, Sprinkle, Berger, de~Heer
  et~al.}}]{orlita:prl08}
\bibinfo{author}{\bibfnamefont{M.}~\bibnamefont{Orlita}},
  \bibinfo{author}{\bibfnamefont{C.}~\bibnamefont{Faugeras}},
  \bibinfo{author}{\bibfnamefont{P.}~\bibnamefont{Plochocka}},
  \bibinfo{author}{\bibfnamefont{P.}~\bibnamefont{Neugebauer}},
  \bibinfo{author}{\bibfnamefont{G.}~\bibnamefont{Martinez}},
  \bibinfo{author}{\bibfnamefont{D.~K.} \bibnamefont{Maude}},
  \bibinfo{author}{\bibfnamefont{A.-L.} \bibnamefont{Barra}},
  \bibinfo{author}{\bibfnamefont{M.}~\bibnamefont{Sprinkle}},
  \bibinfo{author}{\bibfnamefont{C.}~\bibnamefont{Berger}},
  \bibinfo{author}{\bibfnamefont{W.~A.} \bibnamefont{de~Heer}},
  \bibinfo{author}{\bibnamefont{M.}\bibnamefont{ Potemski}}, \bibinfo{journal}{Phys. Rev. Lett.}
  \textbf{\bibinfo{volume}{101}}, \bibinfo{eid}{267601} (\bibinfo{year}{2008}).


\bibitem[{\citenamefont{Novoselov et~al.}(2004)\citenamefont{Novoselov, Geim,
  Morozov, Jiang, Zhang, Dubonos, Grigorieva, and Firsov}}]{novoselov:sci04}
\bibinfo{author}{\bibfnamefont{K.}~\bibnamefont{Novoselov}},
  \bibinfo{author}{\bibfnamefont{A.}~\bibnamefont{Geim}},
  \bibinfo{author}{\bibfnamefont{S.}~\bibnamefont{Morozov}},
  \bibinfo{author}{\bibfnamefont{D.}~\bibnamefont{Jiang}},
  \bibinfo{author}{\bibfnamefont{Y.}~\bibnamefont{Zhang}},
  \bibinfo{author}{\bibfnamefont{S.}~\bibnamefont{Dubonos}},
  \bibinfo{author}{\bibfnamefont{I.}~\bibnamefont{Grigorieva}},
  \bibnamefont{and} \bibinfo{author}{\bibfnamefont{A.}~\bibnamefont{Firsov}},
  \bibinfo{journal}{Science} \textbf{\bibinfo{volume}{306}},
  \bibinfo{pages}{666} (\bibinfo{year}{2004}).

\bibitem[{\citenamefont{Zhang et~al.}(2005)\citenamefont{Zhang, Tan, Stormer,
  and Kim}}]{zhang:nat05}
\bibinfo{author}{\bibfnamefont{Y.}~\bibnamefont{Zhang}},
  \bibinfo{author}{\bibfnamefont{Y.-W.} \bibnamefont{Tan}},
  \bibinfo{author}{\bibfnamefont{H.~L.} \bibnamefont{Stormer}},
  \bibnamefont{and} \bibinfo{author}{\bibfnamefont{P.}~\bibnamefont{Kim}},
  \bibinfo{journal}{Nature} \textbf{\bibinfo{volume}{438}},
  \bibinfo{pages}{201} (\bibinfo{year}{2005}).

\bibitem[{\citenamefont{Miller et~al.}(2009)\citenamefont{Miller, Kubista,
  Rutter, Ruan, de~Heer, First, and Stroscio}}]{miller:sci09}
\bibinfo{author}{\bibfnamefont{D.~L.} \bibnamefont{Miller}},
  \bibinfo{author}{\bibfnamefont{K.~D.} \bibnamefont{Kubista}},
  \bibinfo{author}{\bibfnamefont{G.~M.} \bibnamefont{Rutter}},
  \bibinfo{author}{\bibfnamefont{M.}~\bibnamefont{Ruan}},
  \bibinfo{author}{\bibfnamefont{W.~A.} \bibnamefont{de~Heer}},
  \bibinfo{author}{\bibfnamefont{P.~N.} \bibnamefont{First}}, \bibnamefont{and}
  \bibinfo{author}{\bibfnamefont{J.~A.} \bibnamefont{Stroscio}},
  \bibinfo{journal}{Science} \textbf{\bibinfo{volume}{324}}, \bibinfo{eid}{924}
  (\bibinfo{year}{2009}).

\bibitem[{\citenamefont{Latil et~al.}(2007)\citenamefont{Latil, Meunier, and
  Henrard}}]{latil:prb07}
\bibinfo{author}{\bibfnamefont{S.}~\bibnamefont{Latil}},
  \bibinfo{author}{\bibfnamefont{V.}~\bibnamefont{Meunier}}, \bibnamefont{and}
  \bibinfo{author}{\bibfnamefont{L.}~\bibnamefont{Henrard}},
  \bibinfo{journal}{Phys. Rev. B} \textbf{\bibinfo{volume}{76}},
  \bibinfo{pages}{201402} (\bibinfo{year}{2007}).

\bibitem[{\citenamefont{dos Santos et~al.}(2007)\citenamefont{dos Santos,
  Peres, and Neto}}]{lopes:prl07}
\bibinfo{author}{\bibfnamefont{J.~M. B.~L.} \bibnamefont{dos Santos}},
  \bibinfo{author}{\bibfnamefont{N.~M.~R.} \bibnamefont{Peres}},
  \bibnamefont{and} \bibinfo{author}{\bibfnamefont{A.~H. }
  \bibnamefont{Castro Neto}}, \bibinfo{journal}{Phys. Rev. Lett.}
  \textbf{\bibinfo{volume}{99}}, \bibinfo{eid}{256802} (\bibinfo{year}{2007}).

\bibitem[{\citenamefont{Hass et~al.}(2008)\citenamefont{Hass, Varchon,
  Mill\'{a}n-Otoya, Sprinkle, Sharma, de~Heer, Berger, First, Magaud, and
  Conrad}}]{hass:prl08}
\bibinfo{author}{\bibfnamefont{J.}~\bibnamefont{Hass}},
  \bibinfo{author}{\bibfnamefont{F.}~\bibnamefont{Varchon}},
  \bibinfo{author}{\bibfnamefont{J.~E.} \bibnamefont{Mill\'{a}n-Otoya}},
  \bibinfo{author}{\bibfnamefont{M.}~\bibnamefont{Sprinkle}},
  \bibinfo{author}{\bibfnamefont{N.}~\bibnamefont{Sharma}},
  \bibinfo{author}{\bibfnamefont{W.~A.} \bibnamefont{de~Heer}},
  \bibinfo{author}{\bibfnamefont{C.}~\bibnamefont{Berger}},
  \bibinfo{author}{\bibfnamefont{P.~N.} \bibnamefont{First}},
  \bibinfo{author}{\bibfnamefont{L.}~\bibnamefont{Magaud}}, \bibnamefont{and}
  \bibinfo{author}{\bibfnamefont{E.~H.} \bibnamefont{Conrad}},
  \bibinfo{journal}{Phys. Rev. Lett.} \textbf{\bibinfo{volume}{100}},
  \bibinfo{eid}{125504} (\bibinfo{year}{2008}).


\bibitem[{\citenamefont{Shallcross et~al.}(2008)\citenamefont{Shallcross,
  Sharma, and Pankratov}}]{shallcross:prl08}
\bibinfo{author}{\bibfnamefont{S.}~\bibnamefont{Shallcross}},
  \bibinfo{author}{\bibfnamefont{S.}~\bibnamefont{Sharma}}, \bibnamefont{and}
  \bibinfo{author}{\bibfnamefont{O.~A.} \bibnamefont{Pankratov}},
  \bibinfo{journal}{Phys. Rev. Lett.} \textbf{\bibinfo{volume}{101}},
  \bibinfo{eid}{056803} (\bibinfo{year}{2008}).


\bibitem[{\citenamefont{Shallcross et~al.}(2010)\citenamefont{Shallcross,
  Sharma, Kandelaki, and Pankratov}}]{shallcross:prb10}
\bibinfo{author}{\bibfnamefont{S.}~\bibnamefont{Shallcross}},
  \bibinfo{author}{\bibfnamefont{S.}~\bibnamefont{Sharma}},
  \bibinfo{author}{\bibfnamefont{E.}~\bibnamefont{Kandelaki}},
  \bibnamefont{and} \bibinfo{author}{\bibfnamefont{O.~A.}
  \bibnamefont{Pankratov}}, \bibinfo{journal}{Phys. Rev. B}
  \textbf{\bibinfo{volume}{81}}, \bibinfo{pages}{165105}
  (\bibinfo{year}{2010}).

\bibitem[{\citenamefont{Guy Trambly~de
  Laissardi\`{e}re}(2010)}]{laissardiere:nal10}
\bibinfo{author}{\bibfnamefont{L.~M.} \bibnamefont{Guy Trambly~de
  Laissardi\`{e}re}, \bibfnamefont{Didier~Mayou}}, \bibinfo{journal}{Nano
  Lett.} \textbf{\bibinfo{volume}{10}}, \bibinfo{pages}{804}
  (\bibinfo{year}{2010}).


\bibitem[{\citenamefont{Mele}(2010)}]{mele:prb10}
\bibinfo{author}{\bibfnamefont{E.~J.} \bibnamefont{Mele}},
  \bibinfo{journal}{Phys. Rev. B} \textbf{\bibinfo{volume}{81}},
  \bibinfo{pages}{161405} (\bibinfo{year}{2010}).
  
  
\bibitem[{\citenamefont{Li et~al.}(2010)\citenamefont{Li, Luican, dos Santos,
  Neto, Reina, Kong, and Andrei}}]{andrei:nap10}
\bibinfo{author}{\bibfnamefont{G.}~\bibnamefont{Li}},
  \bibinfo{author}{\bibfnamefont{A.}~\bibnamefont{Luican}},
  \bibinfo{author}{\bibfnamefont{J.~M. B.~L.} \bibnamefont{dos Santos}},
  \bibinfo{author}{\bibfnamefont{A.~H. } \bibnamefont{Castro Neto}},
  \bibinfo{author}{\bibfnamefont{A.}~\bibnamefont{Reina}},
  \bibinfo{author}{\bibfnamefont{J.}~\bibnamefont{Kong}}, \bibnamefont{and}
  \bibinfo{author}{\bibfnamefont{E.}~\bibnamefont{Andrei}},
  \bibinfo{journal}{Nature Physics} \textbf{\bibinfo{volume}{6}},
  \bibinfo{pages}{44} (\bibinfo{year}{2010}).

\bibitem[{\citenamefont{Bistritzer and MacDonald}(2010)}]{bistritzer:prb10}
\bibinfo{author}{\bibfnamefont{R.}~\bibnamefont{Bistritzer}} \bibnamefont{and}
  \bibinfo{author}{\bibfnamefont{A.~H.} \bibnamefont{MacDonald}},
  \bibinfo{journal}{Phys. Rev. B} \textbf{\bibinfo{volume}{81}},
  \bibinfo{pages}{245412} (\bibinfo{year}{2010}).

\bibitem[{\citenamefont{Miller et~al.}(2010)\citenamefont{Miller, Kubista,
  Rutter, Ruan, de~Heer, Kindermann, First, and Stroscio}}]{miller:nap10}
\bibinfo{author}{\bibfnamefont{D.~L.} \bibnamefont{Miller}},
  \bibinfo{author}{\bibfnamefont{K.~D.} \bibnamefont{Kubista}},
  \bibinfo{author}{\bibfnamefont{G.~M.} \bibnamefont{Rutter}},
  \bibinfo{author}{\bibfnamefont{M.}~\bibnamefont{Ruan}},
  \bibinfo{author}{\bibfnamefont{W.~A.} \bibnamefont{de~Heer}},
  \bibinfo{author}{\bibfnamefont{M.}~\bibnamefont{Kindermann}},
  \bibinfo{author}{\bibfnamefont{P.~N.} \bibnamefont{First}}, \bibnamefont{and}
  \bibinfo{author}{\bibfnamefont{J.~A.} \bibnamefont{Stroscio}},
  \bibinfo{journal}{Nature Physics in press}  (\bibinfo{year}{2010}).


\bibitem[{\citenamefont{Dresselhaus and Dresselhaus}(2002)}]{dresselhaus:aip02}
\bibinfo{author}{\bibfnamefont{M.~S.} \bibnamefont{Dresselhaus}}
  \bibnamefont{and}
  \bibinfo{author}{\bibfnamefont{G.}~\bibnamefont{Dresselhaus}},
  \bibinfo{journal}{Advances in Physics} \textbf{\bibinfo{volume}{51}},
  \bibinfo{pages}{1} (\bibinfo{year}{2002}).

\bibitem[{\citenamefont{Neto et~al.}(2009)\citenamefont{Neto, Guinea, Peres,
  Novoselov, and Geim}}]{neto:rmp09}
\bibinfo{author}{\bibfnamefont{A.~H. } \bibnamefont{Castro Neto}},
  \bibinfo{author}{\bibfnamefont{F.}~\bibnamefont{Guinea}},
  \bibinfo{author}{\bibfnamefont{N.~M.~R.} \bibnamefont{Peres}},
  \bibinfo{author}{\bibfnamefont{K.~S.} \bibnamefont{Novoselov}},
  \bibnamefont{and} \bibinfo{author}{\bibfnamefont{A.~K.} \bibnamefont{Geim}},
  \bibinfo{journal}{Rev. Mod. Phys.} \textbf{\bibinfo{volume}{81}},
  \bibinfo{eid}{109} (\bibinfo{year}{2009}).

\bibitem[{\citenamefont{Nilsson et~al.}(2008)\citenamefont{Nilsson,
  Castro~Neto, Guinea, and Peres}}]{nilsson:prb08}
\bibinfo{author}{\bibfnamefont{J.}~\bibnamefont{Nilsson}},
  \bibinfo{author}{\bibfnamefont{A.~H.} \bibnamefont{Castro~Neto}},
  \bibinfo{author}{\bibfnamefont{F.}~\bibnamefont{Guinea}}, \bibnamefont{and}
  \bibinfo{author}{\bibfnamefont{N.~M.~R.} \bibnamefont{Peres}},
  \bibinfo{journal}{Phys. Rev. B} \textbf{\bibinfo{volume}{78}},
  \bibinfo{pages}{045405} (\bibinfo{year}{2008}).

\bibitem[{\citenamefont{Brandt et~al.}(1988)\citenamefont{Brandt, Chudinov, and
  Ponomarev}}]{brandt:bo88}
\bibinfo{author}{\bibfnamefont{N.~B.} \bibnamefont{Brandt}},
  \bibinfo{author}{\bibfnamefont{S.~M.} \bibnamefont{Chudinov}},
  \bibnamefont{and} \bibinfo{author}{\bibfnamefont{Y.~G.}
  \bibnamefont{Ponomarev}}, \emph{\bibinfo{title}{Semimetals I: Graphite and
  its Compounds}} (\bibinfo{publisher}{North-Holland, Amsterdam},
  \bibinfo{year}{1988}).

\bibitem[{\citenamefont{Chung}(2002)}]{chung:jms02}
\bibinfo{author}{\bibfnamefont{D.~D.~L.} \bibnamefont{Chung}},
  \bibinfo{journal}{J. Mater. Sci.} \textbf{\bibinfo{volume}{37}},
  \bibinfo{pages}{1} (\bibinfo{year}{2002}).

\bibitem[{\citenamefont{de~Heer et~al.}(2007)\citenamefont{de~Heer, Berger, Wu,
  First, Conrad, Li, Li, Sprinkle, Hass, Sadowski et~al.}}]{deheer:ssc07}
\bibinfo{author}{\bibfnamefont{W.~A.} \bibnamefont{de~Heer}},
  \bibinfo{author}{\bibfnamefont{C.}~\bibnamefont{Berger}},
  \bibinfo{author}{\bibfnamefont{X.}~\bibnamefont{Wu}},
  \bibinfo{author}{\bibfnamefont{P.~N.} \bibnamefont{First}},
  \bibinfo{author}{\bibfnamefont{E.~H.} \bibnamefont{Conrad}},
  \bibinfo{author}{\bibfnamefont{X.}~\bibnamefont{Li}},
  \bibinfo{author}{\bibfnamefont{T.}~\bibnamefont{Li}},
  \bibinfo{author}{\bibfnamefont{M.}~\bibnamefont{Sprinkle}},
  \bibinfo{author}{\bibfnamefont{J.}~\bibnamefont{Hass}},
  \bibinfo{author}{\bibfnamefont{M.~L.} \bibnamefont{Sadowski}},
  \bibinfo{author}{\bibfnamefont{M. } \bibnamefont{Potemski }},
  \bibinfo{author}{\bibfnamefont{G.} \bibnamefont{Martinez}},
  \bibinfo{journal}{Solid State Comm.}
  \textbf{\bibinfo{volume}{143}}, \bibinfo{eid}{076801} (\bibinfo{year}{2007}).

\bibitem[{\citenamefont{Martin et~al.}(2008)\citenamefont{Martin, Blanter, and
  Morpurgo}}]{martin:prl08}
\bibinfo{author}{\bibfnamefont{I.}~\bibnamefont{Martin}},
  \bibinfo{author}{\bibfnamefont{Y.~M.} \bibnamefont{Blanter}},
  \bibnamefont{and} \bibinfo{author}{\bibfnamefont{A.~F.}
  \bibnamefont{Morpurgo}}, \bibinfo{journal}{Phys. Rev. Lett.}
  \textbf{\bibinfo{volume}{100}}, \bibinfo{eid}{036804} (\bibinfo{year}{2008}).

\bibitem[{\citenamefont{Semenoff et~al.}(2008)\citenamefont{Semenoff, Semenoff,
  and Zhou}}]{semenoff:prl08}
\bibinfo{author}{\bibfnamefont{G.~W.} \bibnamefont{Semenoff}},
  \bibinfo{author}{\bibfnamefont{V.}~\bibnamefont{Semenoff}}, \bibnamefont{and}
  \bibinfo{author}{\bibfnamefont{F.}~\bibnamefont{Zhou}},
  \bibinfo{journal}{Phys. Rev. Lett.} \textbf{\bibinfo{volume}{101}},
  \bibinfo{eid}{087204} (\bibinfo{year}{2008}).

\end{thebibliography}
 \end{document}